\documentclass[num-refs]{wiley-article}

\usepackage{siunitx}
\usepackage{float}
\usepackage{amsmath}
\definecolor{mypink3}{cmyk}{0, 0.7808, 0.4429, 0.1412}
\graphicspath{ {./image/} }
\papertype{Original Article}
\title{A Vector Autoregression Prediction Model for COVID-19 Outbreak}

\author[1]{Qinan Wang}
\author[1]{Yaomu Zhou}
\author[2,3]{Xiaofei Chen}

\affil[1]{Lyle School of Engineering, Southern Methodist University, Dallas, TX, 75275, USA}
\affil[2]{Department of Statistical Science, Southern Methodist University, Dallas, TX, 75275, USA}
\affil[3]{Department of Population \& Data Sciences, UT Southwestern Medical Center, Dallas, TX, 75390, USA}

\corraddress{Xiaofei Chen, PhD\\ Department of Statistical Science, Southern Methodist University, Dallas, TX, 75275, USA}
\corremail{xiaofeic@smu.edu}

\runningauthor{Wang et al.}

\begin{document}

\begin{frontmatter}
\maketitle

\begin{abstract}
Since two people came down a county of north Seattle with positive COVID-19 (coronavirus-19) in 2019, the current total cases in the United States (U.S.) are over 12 million. Predicting the pandemic trend under effective variables is crucial to help find a way to control the epidemic. Based on available literature, we propose a validated Vector Autoregression (VAR) time series model to predict the positive COVID-19 cases. A real data prediction for U.S. is provided based on the U.S. coronavirus data. The key message from our study is that the situation of the pandemic will getting worse if there is no effective control.  

% Please include a maximum of seven keywords
\end{abstract}
\end{frontmatter}

\section{Introduction}
COVID-19 (coronavirus-19) a new type virus, belonging to the Coronaviridae family, spreads from Wuhan, China in 2019.\cite{lu2020genomic} The Coronaviridae family consists of two main subfamilies: Coronavirinae and Torovirinae. These viruses affect the neurological, gastrointestinal, hepatic, and respiratory systems and can be grown by humans, livestock, etc.\cite{chen2020emerging,tang2020estimation,wang2006improved} \\\\
Since the appearance, COVID-19 has infected over 59 million people worldwide. \cite{kufel2020arima}The worst situation experienced by the United States (U.S.) followed by the United Kingdom, Italy, France, and Spain. The U.S. has a cumulative 12 million positive cases up to now. It found itself grappling with the worst outbreak after Italy and Spain.\cite{konarasinghe2020modeling}The Centers for Disease Control and Prevention (CDC) has verified evidence that COVID-19 is distributed from human to human, and has also reported that COVID-19 spreads through touching surfaces, close contact, air, or objects that contain viral particles. In the incubation period, it can spread to others. It should be noted that the incubation period and median age of confirmed cases are 3 days and 47 years respectively. \cite{guan2020clinical,maleki2020time}\\\\ 
The economic and social disruption caused by the pandemic is devastating. The disease prevention and control is eager for a disease prediction guidance. Efficient models for short-term forecasting has a pivotal role to develop strategic planning methods in the public health system. Under the guidance of the prediction model, we know the severity and the trends of epidemic under different strategies. It can arouse public awareness and help government take the most benefit measures to avoid deaths and reduce infection, such as ordered school closure, case-base measures, the banning of public events, the encouragement of social distancing, and lockdown.\\\\
Per literature, a system of differential equations for Susceptible-Infected-Removed (SIR) sequences is a typical mathematical epidemiological model for COVID-19 forecasting.\cite{kufel2020arima,malavika2020forecasting,dhanwant2020forecasting,ndiaye2020analysis,bastos2020modeling,fanelli2020analysis,chen2020time} Joining SIR models,
Khan et al. proposed the SQUIDER compartmental model to predict the coronavirus 2019 spread \cite{khan2020predictive}, and Xu et al. applied the generalized fractional-order SEIR model.\cite{xu2020forecast} The SIR model has a good fitting for the simulation and data of the outbreak in the early stage of the disease. However, the obvious limitations are not limited to that the overall model system has a small external control power, and the number of patients presents a typical exponential growth, which is due to the absence of external drugs and preventive measures.\\\\
Other works on COVID-19 prediction has been carried out in Deep Learning and ARIMA (Auto Regressive Integrated Moving Average) univariate time series model. To assess the dynamics of epidemic diseases, time series analysis tools and deep learning are also widely used in publications. Zeroual et al., Shahid et al., and Chimmula et al. performed the Recurrent Neural Network to predict the spread.\cite{zeroual2020deep,shahid2020predictions,chimmula2020time} With time series tools, Alzahrani et al., Sahai et al., and Kumar et al. predicted the COVID-19 by ARIMA univariate model.\cite{alzahrani2020forecasting,sahai2020arima,kumar2020forecasting} Deep learning requires a high number of training samples. However, the data we have are still few, so the model generalization is unappealing, namely overfitting. In the time series field, the ARIMA model is quite simple, requiring only endogenous variables and no other exogenous variables. A Stationary is required for the time series, or it is stationary after differencing. Essentially, it lays a shortfall in explaining the causality between different variables.\\\\
This article aims to build a generalized VAR (Vector Autoregressive) model for predicting the dynamics of COVID-19 daily cases of the epidemic. VAR is a comprehensive model integrating the advantages of multiple linear regression and the advantages of time series model (the influence of lag term can be analyzed). It applies linear relations to describe a stable system. Under the stationary condition, we can achieve a consistent estimator with the least-square estimation. Besides, VAR can describe the dynamic linear correlation between variables that affect each other, whether used for prediction, interpretation, or sensitivity analysis are clear. With the selected correlated variables among undetected infected, detected deaths, detected recovered, average temperature, precipitation, wind speed, humidity, population density, social trust and civic engagement, that are commonly cited in other epidemiology publications, VAR multivariate model can have a better performance on forecasting and provide an interpretive result. \cite{khan2020predictive,siedner2020social,behnood2020determinants,bialek2020geographic,elgar2020trouble,bruine2020age,james2020cluster}\\\\
The correlated variables we choose are useful for analyzing the critical factors driving epidemics. Not only for COVID-19 but may also this model enlighten other epidemics prediction. For the result, some publications tend to be more concerned with the cumulative positive cases, while this article has a very definite awareness of the daily increase cases. A cumulative positive cases prediction is less meaningful than a daily cases increase, since the latter is a better representative signal for epidemic severity. It is also a critical indicator to access the efficiency of COVID-19 control.\\\\
In the next section (Section 2), we describe the data we used in the analyses. The method section (Section 3) elaborates the proposed VAR model and analysis plan. The Results section (Section 4) provides the prediction results by VAR modeling and an internal validation/evaluation of the model. Section 5 discusses the model performance, further improvement, and comparison with other models.\\\\
\section{The daily reported COVID-19 data}
The COVID-19 disease has been reported by CDC (Centers for Disease Control and Prevention) and published in nation and fifty states in the United States by the Center for Systems Science and Engineering (CSSE) at Johns Hopkins University. We obtain the data from \url{https://covidtracking.com/data/download}  maintained by ``The Atlantic Monthly Group''. The data contain the number of death confirmed, death increasing, death probable, hospitalized, hospitalized cumulative, positive confirmed, positive case viral, positive increasing, etc. The available data is beginning on January 22, 2020 to November 24, 2020 (now). 

\section{Method}
\subsection{A Vector Autoregressive Panel Time Series Model}\label{sec:var}
VAR model was proposed by Christopher Sims in 1980s, using all the current variables in the model to carry out regression for some lagged variables. It is an extension of the AR (autoregression) model, which has been widely used for time series. VAR model takes each endogenous variable as a function of the lag value of all endogenous variables in the system, thus extending the univariate autoregressive model to the "vector" autoregressive model composed of multiple time series variables. \\\\
Let $\mathbf{X}_{t}$ be a causal, stationary multivariate process, then the VAR model can be expressed as:
\begin{equation}\label{multivariate}
    \mathbf{X}_{t}=\boldsymbol{\alpha}+\boldsymbol{\Phi}_{1} \mathbf{X}_{t-1}-\cdots-\boldsymbol{\Phi}_{p} \mathbf{X}_{t-p}+\mathbf{a}_{t}
\end{equation}\\\\
where $\mathbf{X}_{t}=\left(X_{t 1}, \ldots, X_{t m}\right)^{\text{T}}$ is an $m \times t$ matrix; $\boldsymbol{\Phi}_{k}$ is a real-valued $m \times m$ matrix for each $k=1, \ldots, p$; $\mathbf{a}_{t}$ is multivariate white noise with covariance matrix $\mathbb{E}\left[\mathbf{a}_{t} \mathbf{a}_{t}^{\text{T}}\right]=\boldsymbol{\Gamma}_{a}$; 4. $\boldsymbol{\alpha}=\left(\mathbf{I}-\Phi_{1}-\cdots-\Phi_{p}\right) \boldsymbol{\mu},$ and $\boldsymbol{\mu}=\mathbb{E}\left(\mathbf{X}_{t}\right)$;
$\mathbf{I}=\{1,1,\dots,1\}$. Now $\mathbf{X}_{t}$ is called a VAR ($p$) process, that is, a vector AR process of order $p$.\\\\
Equation (\ref{multivariate}) can be expressed in multivariate operator notation way: $\boldsymbol{\Phi}(B)\left(\mathbf{X}_{t}-\boldsymbol{\mu}\right)=\mathbf{a}_{t},$ where
$\boldsymbol{\Phi}(B)=\mathbf{I}-\boldsymbol{\Phi}_{1} B-\cdots-\boldsymbol{\Phi}_{p} B^{p}$ and $B^{k} \mathbf{X}_{t}=\mathbf{X}_{t-k}$.\\\\
A multivariate process $\mathbf{X}_{t}$ satisfying the difference equation in Equation (\ref{multivariate}) is a stationary and causal VAR ($p$) process if and only if the roots of the determinantal equation,$
|\boldsymbol{\Phi}(z)|=\left|\mathbf{I}-\Phi_{1} z-\cdots-\Phi_{p} z^{p}\right|=0$
lie outside the unit circle. A detailed proof see Brockwell et al. and Reinsel et al. \cite{brockwell1991time,reinsel2003elements}\\\\

\subsection{Variables potentially correlated to outcome}
Several potential variables might influence the number of COVID-19 positive cases according to literature \cite{khan2020predictive,siedner2020social,behnood2020determinants,bialek2020geographic,elgar2020trouble,bruine2020age,james2020cluster}: undetected infected, detected deaths, detected recovered, average temperature, precipitation, wind speed, humidity, population density, social trust, civic engagement, that are considered in other publications of epidemic prediction. \\\\
In Chowdhury et al. \cite{chowdhury2018association}, climate changes directly affect five infectious disease transmission. Altered climatic conditions may increase the vector biting rate and the vector's reproduction rate and shorten the pathogen incubation period. Furthermore, depending on the report, If the temperature is higher than 25.0$^{\circ}$C, there is a significant negative correlation between increasing temperature and pneumonia ($p=0.017$). \cite{chowdhury2018association} That is, if the temperature is decreasing under 25.0$^{\circ}$C, pneumonia would spread out faster. \cite{chowdhury2018association} In Liu et al. \cite{liu2014temporal}, when the temperature is lower than 13.0$^{\circ}$C, the number of hospital admission increases, which means the speed of infection also rises up.  Those are the reason why COVID-19 positive confirmed cases appear rebound tendency after October. \cite{liu2014temporal}\\\\
Depending on data reported by the Tasci et al. \cite{tasci2018relationship}, during the periods of high, normal, and low humidity, the number of days admitted with pneumonia was higher at high humidity rates ($p<0.05$). AS a result, the speed of COVID-19 transmission would increase at high humidity situation. In other words, the positive confirmed cases show a significant positive relationship with humidity.\cite{tasci2018relationship}\\\\
According to Brundage et al. \cite{brundage2008deaths}, the pneumonia rate has a stronger positive correlation with mortality. The mortality increase would affect COVID-19 spreading out faster than before. However, the number of death increased would happen after COVID-19 transmission rising. \cite{brundage2008deaths}\\\\
Recovered cases should also have a negative correlation with COVID-19. If recovered cases become more, the number of patients with the virus should be less than before. As a result, fewer patients with the virus would match the lower spread of the virus. When the recovered cases are increasing, the transmission of COVID-19 transmission would be controlled.\\\\

\subsection{Model selection}
As shown in Section~\ref{sec:var}, we need determine the lag order $p$ of the VAR model. 
There are diverse criteria, Akaike Information Criterion (AIC), Hannan-Quinn Criterion (HQC), Schwarz Criterion (SC), and Final Prediction Error (FPE), to find the optimal $p$. 
%-------------------AIC----------------------%
Specifically, AIC is an estimator of out-of-sample prediction error and thereby relative quality of statistical models for a given set of data. \cite{mcelreath2020statistical,taddy2019business} Suppose that we have a statistical model of some data. Let $k$ be the number of estimated parameters in the model and $L$ be the maximum value of the likelihood function for the model. Then the AIC value of the model is $\mathrm{AIC}=2 k-2 \ln (L)$. \cite{burnham2002practical,akaike1974new}
%-------------------HQ----------------------%
HQC is an alternative to AIC and Bayesian information criterion (BIC). It is given as
$\mathrm{HQC}=-2 L +2 k \ln (\ln (n))$,
where $n$ is the number of observations.\\\\
%-------------------SC----------------------%
Schwarz criterion (SC) is given as $\mathrm{SC}=\log (\mathrm{n}) \mathrm{k}-2 \log (\mathrm{L}(\hat{\theta}))$, where $\theta$ is set of all parameter values and $L(\hat{\theta})$ is likelihood of the model returning the data we have, when tested at the maximum likelihood values of $\theta$.  
%-------------------FPE----------------------#
Final Prediction Error (FPE) criterion provides a measure of model quality by simulating the situation where the model is tested on a different data set. It is givin as
$\operatorname{det}\left(\frac{1}{n} \sum_{1}^{n} e\left(t, \hat{\theta}_{i}\right)\left(e\left(t, \hat{\theta}_{i}\right)\right)^\text{T}\right)\left(\frac{1+d_{n}}{1-d_{n}}\right)$, where n is the number of values in the estimation data set, $e(t)$ is a $n$-by-$1$ vector of prediction errors, $\hat{\theta}_{i}$ represents the $i$-th estimated parameters, $d$ is the number of estimated parameters.\\\\
The ordinary least square (OLS) approach is applied to achieve the model estimation. Besides, the model residuals are diagnosed to see if the VAR model assumptions meet.

\section{Result}
\subsection{Preliminary analysis}
\begin{figure}
\centering
\includegraphics[scale=0.35]{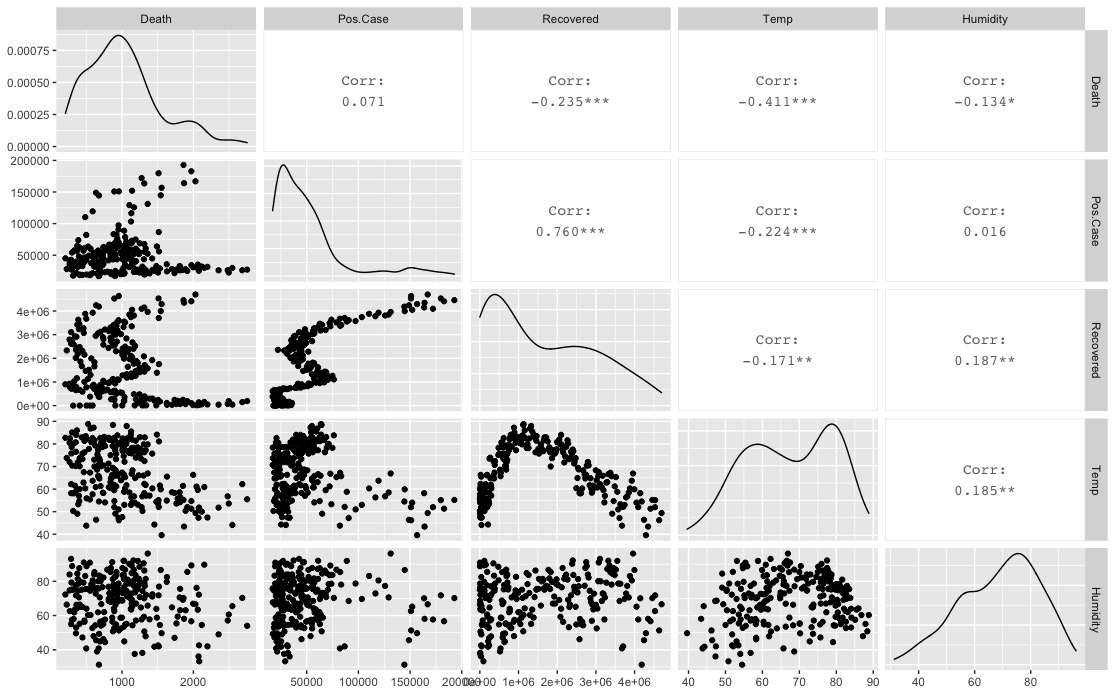}
\caption{Correlation matrix plot. (Death: Cumulative death case; Pos.Case: daily positive case; Recovered: cumulative recovered case; Temp: temperature)}
\label{fig:correlation}
\end{figure}

\begin{figure}
\centering
\includegraphics[scale=0.39]{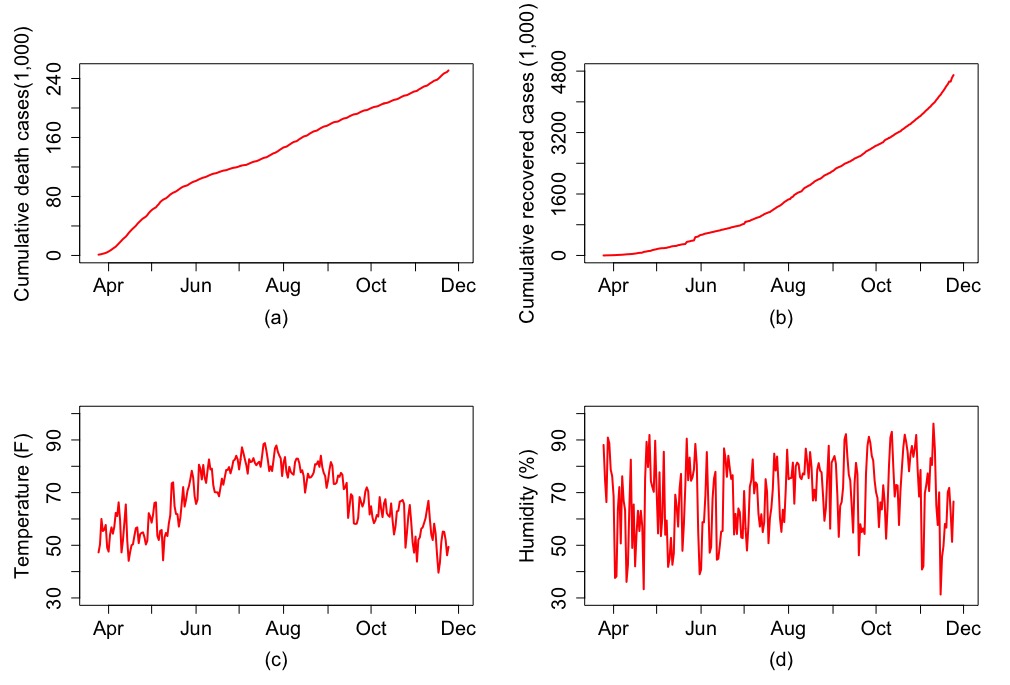}
\caption{Time series from March 25 to November 24, 2020: (a) Cumulative death of COVID-19 cases in US; (b) Cumulative recovered cases; (c) Temperature in Washington D.C.; (d) Humidity in Washington D.C..}
\label{fig:fig1}
\end{figure}

According to literature and correlation analysis, we include cumulative death, cumulative recovered patients, temperature and humidity in the VAR model. Considering the positive cases prediction nationwide, we choose the climate data of Washington D.C. that could be representative. \\\\
The correlation analyses are shown in Figure~\ref{fig:correlation}. Even though 'Death' and 'Humidity' have relatively small correlative coefficients, 0.071 and 0.016, with daily positive case increase, we still keep these two variables. Because the COVID-19 have been verified correlated with cumulative death cases and different humidity.\cite{tasci2018relationship,brundage2008deaths}
\\\\
A descriptive analysis for cumulative death case, cumulative recovered case, temperature, humidity is shown in Figure~\ref{fig:fig1}. The cumulative death cases and the cumulative recovered cases presents straight up tendency.\\\\
Since co-integration between daily positive cases and other selected variables is required by VAR model. We run the co-integration test (Engle Granger test) on all the variables (series). This test is for daily increase positive cases with other variables. The null hypothesis is that there is no co-integration relationship between the two variables. If the variables are all co-integrated with daily positive cases, we can claim that they have are stably correlated in a long run. Results see Appendix Table~\ref{EG}. As all $p$-values are less than 0.05, we have all variables co-integrated with daily increase positive cases. Based on the above results, it is considered that there is a stable relationship and there is no spurious regression for the constructed model. \\\\
%------------------
\\\\
%------------------

\subsection{Model selection}
To determine the lag order of our VAR model, we take AIC, HQC, SC and FPE into consideration (results see Appendix Table~\ref{criteria}). The optimal lag order is determined to be 8. \\\\
With the suggested lag order 8, we estimate our model using ordinary least square technique. We show the parameter estimates in Table~\ref{var}.
\begin{table}
\caption{VAR model parameters estimation$^1$}
\begin{center}
    \begin{tabular}{c|ccccc}
\hline Lag order/ Variables & Death & Pos.Case & Recovered & Temp & Humidity \\
\hline 
       1 & 3.62 & 2.29 & 7.49 & 1.30 & 4.97 \\
       2 & -5.07 & -1.28 & -3.46 & -3.68 & -1.21 \\
       3 & 3.25 & -4.38 & -1.53 & -2.96 & 4.12 \\
       4 & -5.00 & -40.62 & 9.61 & 2.66 & -3.91 \\
       5 & 9.06 & 2.09 & 1.67 & -1.61 &-1.69\\
       6 & 7.43 & -2.46 & -2.92 & -3.96 & 8.07 \\
       7 & -6.89 & 5.80 & -1.50 & 3.61 & -6.47 \\
       8 & 1.53 & -1.29 & 1.94 & -5.22 & -2.83 \\
\hline
\end{tabular}\\
\end{center}
\footnotesize Note: $^1$ ``constant'' item is 8327.24.
\label{var}
\end{table}\\\\
\normalsize
\noindent To verify the assumptions of VAR model, we plot residuals and residuals autocorrelation as shown in Appendix Figure~\ref{fig:var}. The mean value of residual is almost zero (-1.45e-14) and autocorrelation coefficients are within 95\% confidence interval (CI; blue dotted line). We also test the residuals by Ljung-Box test and have p-value 0.20 (null hypothesis is that the data are independently distributed). Hence, the fitted model satisfies the assumptions mentioned above: $\mathbb{E}\left(e_{t}\right)=0$ and $\mathbb{E}\left(e_{t} e_{t-k}^{\text{T}}\right)=0$, where $e_t$ is the residual at time $t$.\\\\

\subsection{COVID-19 daily positive cases prediction}
The objective is predicting the trend of the daily increase positive cases. We predict 30-day daily positive cases starting from July 2, August 21, and November 24, respectively, for internal validation purpose. We pick these three dates for particular reasons. First, 30-day daily increase positive cases after July 2 and August 21 are not fluctuating too much. It is a preliminary test on model performance. Critically, the trend after September 6 becomes steep suddenly. However, the trend before September 6 is similar to before July 2 and before August 21. It may be trapped for the model to identify these three conditions. We want to test if the model will predict the correct rapid increase after September 6.
\begin{figure}
\centering
\includegraphics[scale=0.32]{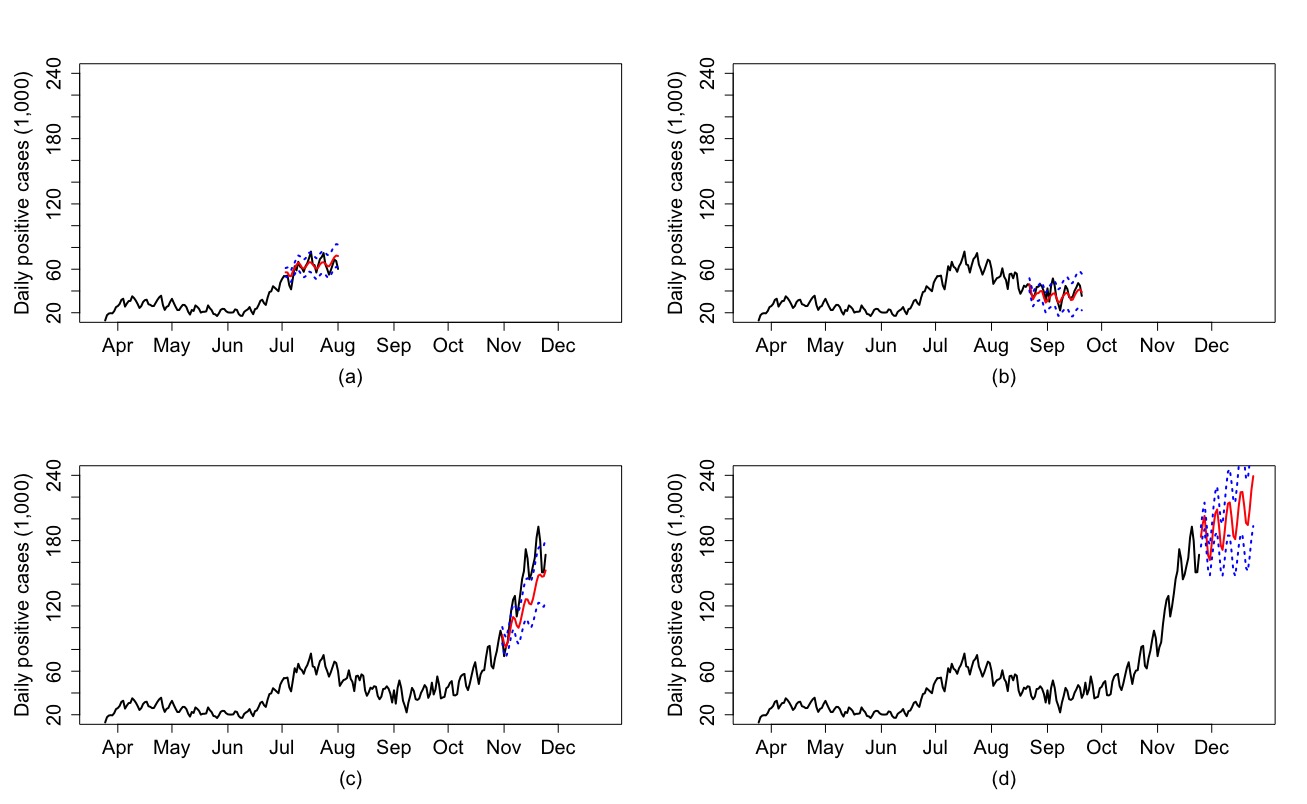}
\caption{Real data (black line) and prediction (red line) with 95\% confidence interval (blue dotted line). \\
(i). For validation purpose: (a) 30 days prediction from July 2, 2020. (b) 30 days prediction from August 21, 2020. (c) 30 days prediction from September 6, 2020. \\
(ii). Real prediction: (d) 30 days prediction from November 24, 2020.}
\label{fig:final}
\end{figure}\\\\
As mentioned, the first three plots (a) -- (c) in Figure~\ref{fig:final} are for internal validation purpose. As we can see that the model is useful, since the real data (black) is covered by the predicted 95\% confidence interval (blue dotted line). To be specific, in Figure~\ref{fig:final}~(a), the black line is the real selected positive confirmed cases daily data, which presents a stable tendency in the first three months, around 40,000 cases every day. After that, in the middle of June, the real data begin to increase. The short red color line is our prediction using proposed model. The black line and the red line are almost overlapped. One thing need to mention is that the black line fluctuates slightly larger than the red, but the predicted is mostly covered by the 95\% confidence interval. It concludes a satisfied prediction. In Figure~\ref{fig:final}~(b), after the middle of July, the number of daily positive confirmed cases decreases and experiences the first peek of 80,000 cases. However, both the black and red lines show a stable trend then, and the figure shows almost the same appearance as the first 30-day prediction. In Figure~\ref{fig:final}~(c), the real data experiences a decreasing trend. But, at the beginning of September, the number of daily positive confirmed cases appears a rebound, directed straight up to the second peak value. The peak value even reaches two hundred thousand cases for one day. Our predicted values are a little lower than real values. The reason can include Halloween holiday parties and some assemblies because those happened at the end of October, and many COVID-19 cases can be confirmed in early November. Furthermore, those are some extrinsic factor besides the ones in our VAR model. As a result, it is reasonable that the black line is higher than the red prediction line and the confidence interval's upper bound. The success is that the model correctly predicts the rapidly increasing trend after August 21.\\\\
The last plot (d) in Figure~\ref{fig:final} is our main result that predicts the daily positive COVID-19 cases 30 days later starting from November 24 (now), that is, a prediction for unknown future trend (to December 24). It is obvious that the the future 30-day growth trend will increase if government are not taking any new measures to control the transmission of COVID-19. During the Christmas, the predicted daily positive case is around 240,000 in US. \\\\
Table~\ref{prediction} shows a comparison of real values and predictions with a 95\% confidence interval. Considering that the daily cases increase data on Monday is partly derived from the cases accumulation over the weekend, we compare the predicted data on each Tuesday with the real values.
\begin{table}
\caption{The real values and predictions of the daily increase positive cases on Tuesday with 95\% confidence interval$^1$.}
\begin{center}
    \begin{tabular}{l|cccc}
\hline Date & Real value & Prediction & Lower 95\% CI & Upper 95\% CI\\
\hline 
       2020-07-09 & 58961 & 64116 & 58240 & 69993  \\
       2020-07-16 & 70446 & 66497 & 57820 & 75173 \\
       2020-07-23 & 71225 & 66129 & 56216 & 76043 \\
       2020-07-30 & 68806 & 71059 & 60634 & 81484 \\
       2020-08-25 & 36588 & 35466 & 28817 & 42115\\
       2020-09-01 & 42426 & 30940 & 21343 & 40536 \\
       2020-09-08 & 22137 & 30037 & 17357 & 42717 \\
       2020-09-15 & 34904 & 32092 & 16668 & 47515 \\
       2020-11-03 & 86662 & 86890 & 77433 & 96348  \\
       2020-11-10 & 131182 & 104960 & 89508 & 120411 \\
       2020-11-17 & 156722 & 121455 & 100033 & 142878 \\
       2020-11-24 & 167012 & 152604 & 123769 & 181438 \\
       2020-12-01 &        & 176995 & 161998 & 191993\\
       2020-12-08 &        & 185548 & 160309 & 210787 \\
       2020-12-15 &        & 194945 & 159893 & 229998 \\
       2020-12-22 &        & 208196 & 164852 & 251539 \\
\hline
\end{tabular}\\
\end{center}
\footnotesize $^1$ Values on December 1, December 8, December 15, December 22 are blank since real data are not available until now.
\label{prediction}
\end{table}
\normalsize
\noindent In Table~\ref{prediction}, the real values are generally within the 95\% confidence level. For the predictions on November 3, 10, 17, 24, the model predicts 86,890, 104,960, 121,455, 152,604 and upper bounds are 96,348, 120,411, 142,878 and 181,438. The real values exceed upper bounds on November 10 and 17 by around 10,000. It still shows the real values are within the confidence interval since the real values do not deviate from the upper bound too far. The model has good performance of catching a rapid increase trend and regular trend.

\section{Discussion}
The study proposed and applied the VAR model for predicting the dynamics of daily COVID-19 positive cases. We selected relevant variables according to literature and checked their correlation coefficients and co-integration. We evaluated our model by comparing the predicted values and real values.\\\\
We can introduce more relevant variables in the future to improve the performance if outside force appears to influence viral transmission, control or exacerbate. The most possible variables available may be the estimation of social distance and the number of vaccination. Then the model will be still useful after vaccine comes out. It enables the model to predict the decrease of infections at the vaccination initial stage. It is also the reason why we investigate the application of VAR model on pandemic predictions. The VAR model is different from and better than SIR and ARIMA. Because SIR and ARIMA have an unsatisfied performance when outside force gets involved. \\\\
The proposed model can be strongly generalized because it is not limited to specific data, since the structure of the model is constructed. Based on the generalization, this model can be used to predict other epidemics with the same characteristics as COVID-19.\\\\

\bibliography{bibliography.bib}

\newpage
\appendix
\section*{Appendix}

\begin{table}[h!]
\caption{Result for Engle-Granger test (co-integration test)}
\label{EG}
\begin{center}
    \begin{tabular}{l|ccc}
\hline Variables & Statistics  & p-value & Co-integration \\
\hline 
       Cumulative death cases & 0.0656 & < .0001 & Y \\
       Cumulative recovered cases & 0.0864 & < .0001 & Y\\
       Temperature & 0.0472 & < .0001 & Y\\
       Humidity & 0.0373  & < .0001 & Y\\
\hline
\end{tabular}
\end{center}
\end{table}

\begin{table}[h!]
\caption{Lag order selection: AIC, HQ, SC, and FPE}
\begin{center}
    \begin{tabular}{c|cccc}
\hline Lag order & AIC  & HQC & SC & FPE \\
\hline 
       1 & 5.57 & 5.59 & 5.62 & 1.57 \\
       2 & 5.50 & 5.54 & 5.59 & 7.82 \\
       3 & 5.48 & 5.53 & 5.61 & 6.52 \\
       4 & 5.47 & 5.53 & 5.63 & 5.67 \\
       5 & 5.45 & 5.53 & 5.65 & 4.76 \\
       6 & 5.44 & 5.54 & 5.68 & 4.45 \\
       7 & 5.42 & 5.52 & 5.69 & 3.42 \\
       8 & 5.39 & 5.51 & 5.69 & 2.49 \\
       9 & 5.39 & 5.53 & 5.73 & 2.55 \\
       10 & 5.39 & 5.55 & 5.78 & 2.76 \\
\hline
\end{tabular}
\end{center}
\label{criteria}
\end{table}

\begin{figure}[H]
\centering
\includegraphics[scale=0.3]{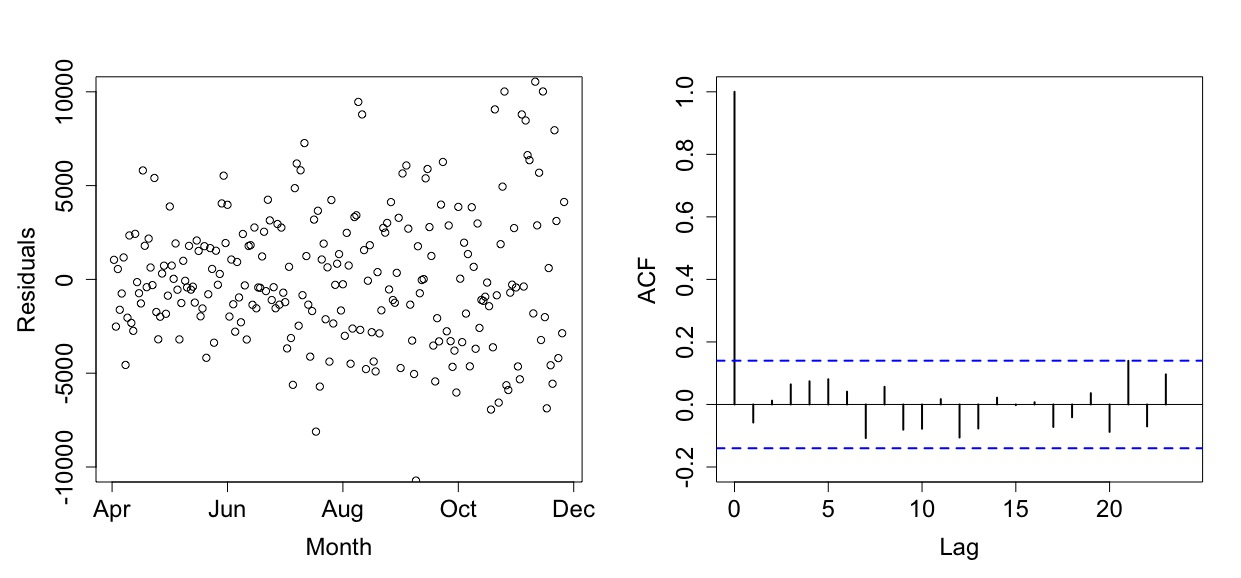}
\caption{Residual plot (left) and its autocorrelation (right). Blue dash lines are upper and lower bound of 95\% confidence interval.}
\label{fig:var}
\end{figure}

\end{document}